% 28th August 1998
\input harvmac.tex
\input epsf.tex
\parindent=0pt
\parskip=5pt

\def\IR{{\hbox{{\rm I}\kern-.2em\hbox{\rm R}}}}
\def\IB{{\hbox{{\rm I}\kern-.2em\hbox{\rm B}}}}
\def\IN{{\hbox{{\rm I}\kern-.2em\hbox{\rm N}}}}
\def\IC{\,\,{\hbox{{\rm I}\kern-.59em\hbox{\bf C}}}}
\def\IZ{{\hbox{{\rm Z}\kern-.4em\hbox{\rm Z}}}}
\def\IP{{\hbox{{\rm I}\kern-.2em\hbox{\rm P}}}}
\def\IH{{\hbox{{\rm I}\kern-.4em\hbox{\rm H}}}}
\def\ID{{\hbox{{\rm I}\kern-.2em\hbox{\rm D}}}}

\noblackbox

\Title{\vbox{\baselineskip12pt
\hbox{hep-th/9808177}}}
{\vbox{\vskip-2cm
\hbox{\centerline{Large $N$ Phases, Gravitational Instantons}}
\smallskip 
\hbox{\centerline{and}} 
\smallskip
\hbox{\centerline{the Nuts and Bolts of AdS Holography}}}}

\centerline{ \bf Andrew Chamblin$^a$,
 Roberto Emparan$^b$, Clifford V. Johnson$^c$ and Robert C.~Myers$^d$ }
\bigskip
\bigskip
\centerline{$^a${\it D.A.M.T.P.,
Silver Street, Cambridge, CB3 9EW, U.K. }}
\medskip
\centerline{$^b${\it Department of Mathematical Sciences,
University of Durham, Durham, DH1 3LE, U.K.}}
\medskip
\centerline{$^c${\it Department of Physics and Astronomy, 
University of Kentucky, Lexington, KY 40506, U.S.A. }}
\medskip
\centerline{$^b${\it Physics Department, McGill
University, Montr\'eal, PQ, H3A 2T8 Canada}}
\footnote{}{\sl email: $^a${\tt H.A.Chamblin@damtp.cam.ac.uk},
$^b${\tt Roberto.Emparan@durham.ac.uk}, $^c${\tt cvj@pa.uky.edu},  
$^d${\tt rcm@hep.physics.mcgill.ca}}
\vskip1.5cm
\centerline{\bf Abstract}
\vskip0.7cm
\vbox{\narrower\baselineskip=12pt\noindent
Recent results in the literature concerning holography indicate that
the thermodynamics of quantum gravity (at least with a negative
cosmological constant) can be modeled by the large $N$ thermodynamics
of quantum field theory. We emphasize that this suggests a completely
unitary evolution of processes in quantum gravity, including black
hole formation and decay; and even more extreme examples involving
topology change.  As concrete examples which show that this
correspondence holds even when the space--time is only {\it locally}
asymptotically AdS, we compute the thermodynamical phase structure of
the AdS--Taub--NUT and AdS--Taub--Bolt spacetimes, and compare them to
a 2+1 dimensional conformal field theory (at large $N$) compactified
on a squashed three sphere, and on the twisted plane.}
\vskip0.5cm

%\draft
\Date{28th August 1998} %Carol's Birthday tomorrow!

\baselineskip13pt

\lref\thooft{ G. 't Hooft, {\it ``Dimensional
Reduction in Quantum Gravity''}, in the proceedings of ``Salamfest,
1993'', gr-qc/9310026.}

\lref\lenny{ L. Susskind, {\it ``The World as a Hologram''}, 
J. Math. Phys. {\bf 36}, 6377 (1995); hep-th/9409089.}

\lref\juan{ J. M. Maldacena, {\it ``The Large $N$ Limit of 
Superconformal Field Theories and Supergravity''}, hep-th/9711200.}

\lref\gub{ S. S. Gubser, I. R. Klebanov and A. M. Polyakov, 
{\it ``Gauge Theory Correlators from Non--critical String Theory''},
hep-th/9802109.  }
\lref\ed{ E. Witten, {\it ``Anti--de Sitter Space and Holography''}, 
hep-th/9802150.  }

\lref\edd{ E. Witten, {\it ``Anti--de Sitter Space, Thermal Phase 
Transition, and Confinement in Gauge Theories''}, hep-th/9803131.}

\lref\gazza{ G. W. Gibbons, {\it ``Wrapping Branes in Space and Time''},
hep-th/9803206.}

\lref\gibbhawk{G. W. Gibbons and S. W. Hawking, 
{\it ``Classification of Gravitational Instanton Symmetries''},
Commun. Math. Phys. {\bf 66} (1979) 291.}

\lref\kramer{D. Kramer, E. Herlt, M. MacCallum and H. Stephani,
 {\it ``Exact Solutions of Einstein's Field Equations''},
ed. E. Schmutzer, Cambridge University Press, 1979.}

\lref\gross{D. J. Gross and E. Witten, {\it 
``Possible Third Order Phase Transition in the Large $N$ Lattice Gauge
Theory''}, Phys. Rev. {\bf D21} (1980) 446.}

\lref\taubnut{A. H. Taub, {\it ``Empty 
Space--Times Admitting a Three Parameter Group of Motions''}, Annal.
Math. {\bf 53} (1951) 472\semi E. Newman, L. Tamborino and T. Unti,
J. Math. Phys. {\bf 4} (1963) 915.}

\lref\gibbs{G. Gibbons and S. W. Hawking, 
{\it ``Action Integrals and Partition 
Functions in Quantum Gravity''}, Phys. Rev. {\bf D15} (1977) 2752.}

\lref\hawkhoro{S. W. 
Hawking and G. Horowitz, {\it ``The Gravitational Action, Entropy and
Surface Terms''}, Class. Quant. Grav. {\bf 13} (1996) 1487,
gr-qc/9501014.}

\lref\sjrey{S--J. Rey, {\it ``Holography Principle and Topology 
Change in String Theory''},  hep-th/9807241.}

\lref\hawkpope{S. W. Hawking and C. N. Pope,
 {\it ``Generalized Spin Structures in Quantum Gravity''},
Phys. Lett. {\bf 73B} (1978) 42.}

\lref\adm{R. Arnowitt, S. Deser and C. Misner, in {\it ``Gravitation: 
An Introduction to Current Research''}, ed. L. Witten, Wiley, New York
(1962).}

\lref\misner{C. Misner, {\it ``The Flatter 
Regions of Newman, Unti and Tamburino's 
Generalized Schwarzschild Space''}, Jour. Math. Phys. {\bf 4} (1963)
924.}

\lref\beckhawk{J. Bekenstein, {\it ``Black 
Holes and Entropy''}, Phys. Rev. {\bf D7} (1973) 2333\semi
S. W. Hawking, {\it ``Particle Creation by Black Holes''},
Commun. Math. Phys. {\bf 43} (1975)~199.}

\lref\don{D. N. Page, {\it ``Taub--NUT Instanton with a Horizon''},
Phys. Lett. {\bf 78B}, 249-251 (1978).}

\lref\hawkpage{S. W. Hawking and D. N. Page, {\it ``Thermodynamics of 
Black Holes in Anti--de Sitter Space''}, Comm. Math. Phys. {\bf 87},
577 (1983).}

\lref\hawkhunt{S. W. Hawking and C. J. Hunter,
{\it ``The Gravitational Hamiltonian in the Presence of 
Non--Orthogonal Boundaries''},
DAMTP/R-96/9, gr-qc/9603050.}

\lref\hunt{C. J. Hunter, {\it ``The Action of Instantons with 
Nut Charge''}, gr-qc/9807010.}

\lref\virtual{ S. W. Hawking, {\it ``Virtual Black Holes''},
Phys. Rev. {\bf D53}, 3099-3107, (1996); hep-th/9510029.}

\lref\simon{S. W. Hawking and S. F. Ross,
 {\it ``Loss of Quantum Coherence through Scattering off Virtual Black
Holes''}, Phys. Rev. {\bf D56} (1997) 6403,  hep-th/9705147.}

\lref\gazzorigin{ G. W. Gibbons, in {\it ``Fields and Geometry''},
Proceedings of the 22nd Karpacz Winter School of Theoretical Physics,
Karpacz, Poland, 1986, edited by A. Jadczyk, World Scientific,
Singapore, (1986).}

\lref\robb{R. B. Mann, 
{\it ``Charged Topological Black Hole Pair Creation''},
Nucl. Phys. {\bf B516}, 357-381, (1998); hep-th/9705223.}

\lref\igor{I. R. Klebanov and  A. A. Tseytlin,
 {\it ``Entropy of Near--Extremal Black $p$--Branes''},
Nucl. Phys. {\bf B475} (1996) 164, hep-th/9604089.}

\lref\seiberg{N. Seiberg, {\it ``Notes on Theories with Sixteen 
Supercharges''}, Prog. Theor. Phys. Suppl. {\bf 102} (1990) 319,
 hep-th/9705117\semi S. Sethi and L. Susskind, {\it ``Rotational
 Invariance in the M(atrix) Formulation of Type IIB Theory''},
 Phys.Lett. B400 (1997) 265, hep-th/9702101.}

\lref\topo{ See, {\it e.g.,} R. B. Mann,
 {\it ``Pair Production of Topological anti de Sitter Black Holes''},
Class. Quant. Grav. {\bf 14} (1997) L109, gr-qc/9607071\semi D. Brill,
J. Louko, P. Peld\'an, {\it ``Thermodynamics of (3+1)--dimensional
black holes with toroidal or higher genus horizons''}, Phys. Rev. {\bf
D56} (1997) 3600, gr-qc/9705012\semi L. Vanzo, {\it ``Black holes with
unusual topology''}, Phys. Rev. D56 (1997) 6475, gr-qc/9705004.}

\lref\roberto{R. Emparan, {\it ``AdS Membranes Wrapped on Surfaces of 
Arbitrary Genus''}, Phys. Lett. {\bf B432} (1998) 74, hep-th/9804031.}

%%%%%%%%%%%%%%%%%%%%%%%%%%%%%%%%%%%%%%%%%%%%%%%%%%%%%%%%%%%%%%%%%%%%%%%%%%

``To them, I said, 
the truth would be literally nothing but the shadows of the images.''
\hfill
 \rightline{\it --Plato, The Republic (Book VII)}

\newsec{Introduction and Motivation}

The holographic principle\refs{\thooft,\lenny}\ asserts that all of
the information contained in some region of space--time may be
represented as a ``hologram'': a theory which lives on the boundary of
the region.  The principle also requires that the theory on the
boundary should contain at most one degree of freedom per Planck area.
It follows from these two simple assumptions that the maximum number
of quantum degrees of freedom, which can be stored in a region bounded
by a surface of area $A$, will never exceed $\exp({A\over 4G})$ (where
$G$ is Newton's constant).  This dovetails nicely with the laws of
black hole thermodynamics (which provided some of the inspiration for
the holographic principle), leading some investigators to conclude
that the holographic principle may be an essential ingredient in the
construction of a complete quantum theory of gravity.

Recently, it has been conjectured\refs{\juan,\gub,\ed}\ that
information about the physics of superconformal  field theories
(in the large $N$ limit\foot{Here, ``$N$'' refers to that of $U(N)$
gauge theory in the simplest case of $p{=}3$, with suitable
generalizations in the other cases of $p$.}) may be obtained by
studying the region near the horizon of certain $p$--branes, which
yields a gauged supergravity compactification involving $p{+}2$
dimensional Anti--de Sitter space--time, denoted AdS$_{p+2}$.  The
correspondence is holographic\ed\ because the conformal field
theory (CFT) lives on the causal boundary of AdS.  This boundary is
the ``horosphere'' at infinity\gazza\ --- it is a timelike
hypersurface with the topology $S^{1}{\times}S^{p}$, where the circle
$S^1$ is the (Euclideanized) timelike factor.

The key feature of this AdS--CFT correspondence is the fact that
fields propagating in the bulk of AdS are {\it uniquely} specified by
their behaviour at the boundary.  This allows one to calculate
correlation functions in the boundary theory by calculating the
effective action in the bulk for field configurations which
asymptotically approach the given boundary data\ed.

Given this correspondence, one is naturally led to consider bulk
supergravity spacetimes which are asymptotically equivalent to AdS.
Since the AdS--CFT correspondence asserts that the generating
functional of (large $N$) superconformal field theory propagators on
the boundary, $M$, of AdS are equivalent to supergravity partition
functions in the bulk, it is of some interest to understand how many
such distinct bulk manifolds, $B_i$, with boundary~$M$, may exist.

A more complete version of the conjecture states that the full $1/N$
expansion of the field theory partition function, $Z_{CFT}(M)$, on
$M$, must be expressed as a sum over the $B_i$:
\eqn\sumup{Z_{CFT}(M) = \sum_{i} Z(B_i)
} where $Z(B_i)$ is the {\it string theory} (or {\it M--theory}) partition
function on $B_i$. The stringy part of the story controls the short
distance bulk physics (where gravity alone would fail). In the
stricter large $N$ limit, the string theory reduces to gravity, valid
on spacetimes of low curvature (whose typical length scale, $l$, is of
the order $N^{f(p)}$, where $f(p)$ is some positive function of $p$),
and this is the regime we will focus on in this paper.

Recently\refs{\ed,\edd}\ this relation has been employed to study the
large $N$ thermodynamics\foot{Of course, there is a thermodynamic
limit even in finite volume if we take the number of degrees of
freedom, here measured by some power of $N$, to infinity. So we may
indeed have phase transitions\gross.}\ of conformal field theories
(defined at finite temperature by Euclideanizing to periodic time) on
the boundary $S^1{\times}S^p$. (Here, $S^1$ is the Euclidean time.)
There are two known (asymptotically AdS) bulk solutions with this
boundary. The more obvious one is AdS itself (with suitable
identifications), while the other is the Euclidean AdS--Schwarzschild
solution. It was shown that the former solution governs the low
temperature phase of the boundary conformal field theory while the
latter controls the high temperature phase. Many qualitative features
of the dynamics of the finite temperature field theory were reproduced
with these spacetimes, including the geometric behaviour of spatial
and temporal Wilson lines, confirming that the high and low
temperature phases have distinct physical characteristics.  This is a
dramatic demonstration of the properties (and uses of) a holographic
relationship or ``duality'' between two theories.

We would like to emphasize that the arrow runs both ways in this
relationship. While the existence of ---and transition between--- two
different phases of a field theory are uncontroversial concepts to
most theorists, this is not the same for many processes in quantum
gravity. Indeed, as many of the transitions between different
space--time solutions in gravity are not completely understood, there
is still room to assume that ---especially in the cases involving the
evaporation or formation of black holes--- the quantum processes may
be non--unitary. It is also of considerable technical interest as to
how to describe completely such processes, as they often describe
spacetime topology change to relate the initial and final
states.

Crucially, note that in having a holographic relation between field
theory and gravity (at least with negative cosmological constant), we
have a powerful laboratory for studying those bulk topology change
processes which are still a matter of debate\foot{See ref.\sjrey\ for
a recent discussion ---with a different flavour--- of space--time
topology change in this context.}. In particular, the relation to
field theory (if proven) completely removes the possibility of a
non--unitary nature of the processes governing spacetime topology
change in quantum gravity with negative cosmological constant, and we
find it highly suggestive of a similar conclusion for all
gravitational situations.

In the field theory examples of ref.\refs{\ed,\edd}, (specializing to
the case $p{=}2$), while the boundary field theory phase transition
takes place, the dominant contribution on the right hand side of
\sumup\ 
shifts from AdS$_4$, with topology $\IR^3{\times}S^1$, to
AdS$_4$--Schwarzschild, with topology $\IR^2{\times}S^2$. This
transition was studied originally in ref.\hawkpage.  The nature of
this phase transition is intimately associated with the fact that the
gravitational potential of AdS behaves more or less like a large,
perfectly insulating ``box''.  Massive particles are confined to the
interior of AdS, and while massless modes may escape to infinity, the
fluxes for incoming and outgoing radiation in a thermal state at
infinity are equal (the causal boundary acts like a mirror).

It was shown in ref.\hawkpage\ that there is a critical temperature,
$T_c$, past which thermal radiation is unstable to the formation of a
Schwarzschild black hole.  (In fact, they found that for $T{>}T_c$
there are two values of the black hole mass at which the Hawking
radiation can be in equilibrium with the thermal radiation of the
background.  The lesser of these two masses is a point of unstable
equilibrium (it has negative specific heat), whereas the greater mass
is a point of stable equilibrium.)

Since a phase transition in the field theory is a unitary process,
this means that it would seem that there is no ``information loss'',
or loss of unitarity, in the bulk physics involving the nucleation and
evaporation of black holes as one moves between the various phases.
It would be certainly interesting to see if this unitary conformal
field theory description extends to other transitions between
instantons which involve space--time topology change. Clearly, this
would then be in sharp contrast to the claims of recent
authors\refs{\virtual,\simon}, who have argued that whenever there is
a topology changing transition ({\it i.e.}, by black hole pair
creation or some other process), the superscattering matrix will not
factorize into and S--matrix and its adjoint and hence there will be a
loss of quantum coherence.
%of fat pookies.

It would therefore seem, at first glance, that the AdS version of the
holographic principle has provided us with a {\it precise} argument
which shows that information is not lost in black hole evolution or
topology changing transitions, at least as long as the topology change
occurs in a spacetime which is asymptotically AdS.

This suggests an interesting and vigorous program of revisiting the
study of various spacetime transitions between many instantons of
interest, now in an AdS context.

In this note, we will extend the holography laboratory to include
examples with non--trivial topology, and  which are only {\it locally}
asymptotically AdS.  We discuss the Taub--NUT--AdS (TN--AdS) and
Taub--Bolt--AdS (TB--AdS) space--times.  These space--times have a
global non--trivial topology due to the fact that one of the Killing
vectors has a zero--dimensional fixed point set (``nut'') or a
two--dimensional fixed point set (``bolt'').  Further, these
four--dimensional space--times have Euclidean sections which cannot be
exactly matched to AdS at infinity.

We show that it is possible to have a thermally triggered phase
transition from TN--AdS to TB--AdS, which is the natural
generalization of the Hawking--Page phase transition from AdS to
Schwarzschild--AdS. We also notice that in the limits where we can use
the naive field theory expectations, the results are in agreement with
boundary field theory.

In the first case under study, where the bolt is an $S^2$, the
presence of these nuts or bolts implies that the bulk supports a
non--trivial NUT--charge, which in turn implies that the boundary must
be realized as an $S^1$ bundle over $S^2$ ({\it i.e.}, the Chern
number of this Hopf fibration (denoted $C_1$) is related to the NUT
charge $N$ in the bulk by the explicit relation\hunt\ $N{=}{1\over
8\pi}{\beta}C_1$, where $\beta$ is the period of the $S^1$ fibre at
infinity); the boundary at infinity is a ``squashed'' three--sphere.

This squashed three--sphere is the three dimensional space on which
the boundary conformal field theory will be compactified, with $\beta$
identified with the inverse temperature, in analogy with the
AdS/AdS--Schwarzschild system\hawkpage. As studied in
refs.\refs{\ed,\edd}, we see that the bulk behaviour is consistent
with the expected phase structure of the conformal field theory on the
boundary.

In the second case, the bolt is an $\IR^2$, and the resulting absence
of a non--trivial fibration means that there is no link between the
temperature at infinity and the squashing parameter. The squashing
parameter describes a fixed deformation of the boundary as a twisted
product of $\IR^2$ and Euclidean time $S^1$. In this case, the phase
structure found in the bulk again is consistent with that of conformal
field theory on the boundary.

\newsec{The NUTs and Bolts of AdS}

We now turn our attention to a particular class of metrics which are
locally asymptotically equivalent to anti--de Sitter space--time: the
Taub--NUT--AdS (TN--AdS) and Taub--Bolt--AdS (TB--AdS) metrics.  The
metric on the Euclidean section of this family of solutions may be
written in the form\kramer\
\eqn\family{ds^2 = V(r) (d\tau + 2n\cos\theta d\varphi)^2 
+ {dr^2\over V(r)} + (r^2 -n^2)(d\theta^2+\sin^2\theta d\varphi^2)}
where
\eqn\where{V 
= { (r^2 +n^2)- 2 mr +l^{-2}( r^4 - 6 n^2 r^2-3 n^4)\over r^2 - n^2}}
and we are working with the usual convention ($l^2{=}-\Lambda /3$),
with $\Lambda{<}0$ being the cosmological constant.  Here, $m$ is a
(generalized) mass parameter and $r$ is a radial coordinate.  Also,
$\tau$, the analytically continued time, parameterizes a circle,
$S^1$, which is fibred over the two sphere $S^2$, with coordinates
$\theta$ and $\varphi$.  The non--trivial fibration is a result of a
non--vanishing ``nut parameter'' $n$.

In the asymptotic region, the metric \family\ becomes
\eqn\becomes{ds^2={l^2\over r^2}dr^2 +r^2\left[
{4n^2\over l^2}\left( d\psi+\cos\theta d\varphi\right)^2 +
d\theta^2+\sin^2\theta d\varphi^2
\right]}
where $\psi{=}\tau/2n$. One can recognize the angular part of the
metric as that of a ``squashed'' three--sphere, where $4n^2/l^2$
parameterizes the squashing.  This finite amount of squashing
contrasts with the standard Taub--NUT solution\taubnut\ with
$\lambda{=}0$. In the latter, a squashed three--sphere also arises in
the asymptotic region, but $4n^2/l^2$ is replaced by $4n^2/r^2$ in the
angular part of the metric ({\it c.f.} \becomes).

Remarkably, this asymptotic metric
\becomes\ is still maximally symmetric, to leading order, {\it i.e.,}
$R_{\mu\nu\alpha\beta}{=}-1/l^2(g_{\mu\alpha}g_{\nu\beta}-
g_{\mu\beta}g_{\nu\alpha})$. Hence
we can still think of these solutions as locally asymptotically
AdS$_4$.

\subsec{Taub--NUT--AdS}

To begin with, let us restrict our attention to nuts, the zero
dimensional fixed point set.  For a regular nut to exist we need to
satisfy the following conditions:

\noindent (a) 
In order to ensure that the fixed point set is zero dimensional, it is
necessary that the Killing vector ${\partial}_{\tau}$ has a fixed
point which occurs precisely when the ($\theta,\varphi$) two-sphere
degenerates, {\it i.e.}, $V(r{=}n){=}0$.

\noindent (b) In order for the ``Dirac--Misner''\misner\
 string to be unobservable, it is necessary that the period of $\tau$
satisfy $\Delta\tau{=} 4n\Delta\varphi$. Since we want to avoid conical
singularities at the poles of the angular spheres, then
$\Delta\varphi{=}2\pi$, and therefore $\Delta\tau{=}8\pi n$.

\noindent (c) In general, these constraints will make 
the point $r{=}n$ look like the origin of~${\bf R}^4$ with a conical
deficit. In order to avoid a conical singularity, the fiber has to
close smoothly at $r{=}n$. This requires $\Delta\tau V'(r=n) =4\pi $,
{\it i.e.}, $V'(r=n) ={1/2n}$.

Now, condition (a) requires that the numerator of $V$ has a double
zero at $r{=}n$.  It is easy to see then that the ``mass'' parameter
$m$ must be:
\eqn\massone{m_{n}=n-{4n^3\over l^2} }
and then
\eqn\then{V_n(r)= {r-n \over r+n} [1 +l^{-2} (r-n)(r +3n)].
} With this, condition (c) is automatically satisfied. This is due to
the fact that the term that multiplies the cosmological constant
vanishes at the nut, and what remains is the same as in the familiar
case with $\Lambda{=}0$. This means that the presence of a cosmological
constant does not affect the nut.

It is interesting to notice that here with $\Lambda{<}0$, $m$ does not
need to be positive in order for the nut to be regular. It is also
worth remarking that $n$ remains an arbitrary parameter, which will be
assumed to be positive, without loss of generality. That is, as $n$
varies in this family, we see that the squashing of the asymptotic
three--spheres changes, and thus for fixed cosmological constant we
have a one--parameter family of TN--AdS solutions.

Note that for the special case $n{=}l/2$ the squashing in \becomes\
vanishes, {\it i.e.}, the asymptotic spheres are round. In fact, in
this special case, the geometry coincides precisely with the AdS$_4$
space.  In order to see this, change $\tau$ to the more usual $\psi$
coordinate in $S^3$, $\tau{=}2 n\psi$, so that the period of $\psi$ is
$4\pi$.  It is convenient to perform another coordinate change
on~\family, by shifting $r{\rightarrow}r{+}n$ to find
\eqn\new{\eqalign{ds^2 &= {U(r)\over f(r)}dr^2+ 4n^2 {f(r)\over U(r)} 
(d\psi + \cos\theta d\varphi)^2 +r^2 U(r) (d\theta^2 +\sin^2\theta
d\varphi^2),\cr {\rm with}\,\, f(r)&=1+{r^2\over l^2}\left(1+{4n\over
r}\right)\cr {\rm and}\,\, U(r)&=1+{2 n\over r} }} The nut is now at
$r=0$.

Now start from the following form for the metric on AdS$_4$ as the
Poincar\'e ball\ed:
\eqn\ball{ds^2=4{dy^2 +y^2 d\Omega_3^2 \over (1-y^2/l^2)^2}}
The boundary is at $y{=}l$, and it is an $S^3$. Changing coordinates
according to
\eqn\change{{y^2\over l^2}={r\over r+l}}
so that the boundary is now at $r{\rightarrow}\infty$, we find that
the following metric for AdS$_4$:
\eqn\following{ ds^2={l^2\over r^2}\left({dr^2\over 1+l/r}\right)
+r^2\left(1+{l \over r}\right) [(d\psi+ \cos\theta d\varphi)^2 +
d\theta^2 +\sin^2\theta d\varphi^2].} This AdS metric coincides
precisely with the TN--AdS metric \new\ with $n{=}l/2$. At $r{=}0$
there is a coordinate singularity, but this is easily seen to be just
like the origin of~$\IR^4$, {\it i.e.}, a nut.  It is not surprising
to find a slicing where AdS$_4$ contains a nut: given any point in the
Poincar\'e ball, we can always choose coordinates such that it looks
like the origin of~$\IR^4$.

One can confirm that in general the TN--AdS metric is distinct from
AdS$_4$ by comparing curvature invariants, {\it e.g.}
$R^{\mu\nu}R_{\mu\nu}$, on the two spaces.

\subsec{Taub--Bolt--AdS}

We begin by casting the metric (2.1) in the form
\eqn\cast{ds^2 = 4n^2 V(r) (d\psi +
  \cos\theta d\varphi)^2 + {dr^2\over V(r)} + 
(r^2 -n^2)(d\theta^2 +\sin^2\theta d\varphi^2)}
with 
\eqn\anotherwith{V_b(r)
 ={r^2 -2mr +n^2 +l^{-2} (r^4 -6n^2 r^2 - 3n^4)  \over r^2-n^2}.}
where as usual $\psi$ has period $4\pi$.  In order to have a regular
bolt at $r{=}r_b{>}n$ the following conditions must be met:

\noindent (a) $V(r_b)=0.$

\noindent (b) $V'(r_b)={1/2n}.$

These are rather like the conditions for having a nut, but since 
$r_b{>}n$, the fixed point set of $\partial_\psi$ is two dimensional,
instead of zero dimensional.  Moreover, the zero of the numerator of
$V(r)$ at $r{=}r_b$ must now be a single one.

After some simple algebra, we find that condition (a) imposes
\eqn\implies{m=m_{b}={r_b^2+n^2 \over 2 r_b} 
+{1\over 2 l^2} \left(r_b^3 -6n^2 r_b -3{n^4 \over r_b}\right).}  Then
we find
\eqn\findit{V'(r_b)=
{3 \over l^2}\left({r_b^2 -n^2 +l^2/3\over r_b}\right).}
Now we require (b) to be satisfied. The ensuing equation yields $r_b$
as a function of $n$ and~$l$:
\eqn\boltplace{r_{b\pm}=
{l^2 \over 12n} \left(1\pm\sqrt{1-48{n^2\over l^2}+144 {n^4 \over
l^4}}\right).}  For $r_b$ to be real the discriminant must be
non--negative. Futhermore we must take the part of the solution which
corresponds to $r_b{>}n$. This gives:
\eqn\real{n\leq\left({1\over 6}-{\sqrt{3}\over 12}\right)^{1\over2}l
=n_{\rm max}.}  {\it It is only for this range of
parameters that one can construct real Euclidean TB--AdS solutions.}
Notice, in particular, that the AdS value $l{=}2n$ lies outside this
range.

It is worth noting that the properties of Taub--bolt in AdS (for the
upper branch, $r_{b+}$) are very different from those of Taub--bolt in
an asymptotically locally flat (ALF) space. The reason is that these
upper branch TB--AdS solutions do not go smoothly onto ALF--TB as the
cosmological constant is switched off.  As $l$ is taken to infinity,
we can see that $r_{b+}{\rightarrow}\infty$. The ALF--TB limit can be
achieved only with the $r_{b-}$ branch TB--AdS solutions. In those
cases, $r_{b-}{\rightarrow}2n$ as the cosmological constant goes to
zero, reproducing the ALF--TB value.

The lower branch family is more analogous to the Schwarzschild--AdS
solutions. In the latter, when the bolt (the Euclidean horizon) is
much smaller than the AdS scale, it resembles closely the
corresponding asymptotically flat bolt. It is only when the black hole
grows enough in size that the AdS structure shows up. By contrast, for
the upper brach TB--AdS solutions, the fact that they live in Anti--de
Sitter space is always relevant.

Interestingly, the global topology of the TB--AdS solution is quite
unlike that of TN--AdS.  Arguments similar to those put forward in
ref.\don\ lead to the conclusion that this solution has the topology
of ${\IC}{\IP}^2{-}\{0\}$, where the removed ``point'' $\{0\}$
corresponds to the squashed three--sphere at infinity.  Furthermore,
the bolt itself may be interpreted as the two--cycle in ${\IC}{\IP}^2$
with odd self--intersection number, {\it i.e.}, this space--time does not
admit any spin structure\foot{This would suggest that there might be
problems with interpreting this as a supergravity
compactification. Recall however, that there is the possibility of
introducing a generalized spin structure\hawkpope, particularly in the
case of $\IC\IP^2$. Even without that possibility, we expect that
holography in AdS$_4$ (and related spacetimes) is a property which
exists independently of the possibility of supergravity
compactifications.}.

Now that we have understood the structure of the TN--AdS and TB--AdS
solutions, we need to examine the possibility of transitions between
them. In order to understand the conditions for this phase transition,
we need to calculate the actions for TN--AdS and TB--AdS.

\subsec{The Action Calculation}

The Euclidean action is given by the formula\gibbs\hawkhoro:
\eqn\formula{I = -{1\over 16{\pi}G} 
\int_{\cal M} d^4x\, \sqrt{g}(R - 2{\Lambda})
- {1\over 8\pi} \int_{\partial \cal M} d^3x\,\sqrt{\gamma}\Theta,}
where $\cal M$ is a compact region of the spacetime, with boundary
$\partial \cal M$ (which we will ultimately send to infinity). Here,
$\gamma_{\mu\nu}$ is the induced metric on $\partial \cal M$, and
$\Theta$ is the trace of the extrinsic curvature of $\partial \cal M$
in $\cal M$. Of course, both of the terms above diverge as the
boundary goes to infinity. Hence to produce a finite and well--defined
action as the boundary $\partial
\cal M$ goes to infinity,
 we will subtract an infinite contribution from a background or
reference space--time solution.  

For a background to be suitable for a given spacetime whose action we
wish to compute, we must match the metric that it
induces on  $\partial \cal M$ to the metric induced
by the spacetime on $\partial
\cal M$, to an order that is sufficient to ensure that the difference
disappears in the limit where we take $\partial\cal M$ to infinity.
Here, this does not seem to be possible using AdS$_4$ as a reference
solution. However, given the asymptotic structure of the TN--AdS and
TB--AdS instantons, it is natural to use TN--AdS as the background for
the solutions described by the metric \family\ which have the same
asymptotic behaviour.  It follows that the action of TN--AdS is
defined to be {\it zero}, because it is regarded as the ground state.

The calculation of the action of TB--AdS relative to TN--AdS is just
the ``nutty'' generalization of the calculations \refs{\hawkpage,\edd}
of the action of AdS--Schwarzschild relative to AdS.  Just as with
these previous calculations, the surface term in \formula\ does not
make any contribution.  It follows that we just need to focus on the
bulk contribution.  Since we are in four dimensions, and working with
solutions of the vacuum Einstein equations, it follows that the Ricci
scalar is given as $R = 4{\Lambda}$, whence the bulk action term
assumes the form
\eqn\assumes{I = -{\Lambda\over 8{\pi}G} \int_{\cal M}\! d^4x\, 
\sqrt{g} = {3\over 8{\pi}Gl^2} {\rm Vol}(M).}

We now need to compare the infinite volume contribution of TB--AdS to
the infinite contribution of TN--AdS; this difference should give us a
finite, physically meaningful answer.  For both metrics, one
calculates the determinant as
\eqn\determinant{\sqrt{g} = 2n(r^2 - n^2)\sin{\theta}}
Taking as our hypersurface $\partial\cal M$ the fixed radius surface
$r{=}R$, the volume contributions from TB--AdS  and TN--AdS
 thus take the explicit form
\eqn\explicit{{\rm Vol}_{b}(R)
 = 2n\int_{0}^{4\pi}\!d{\psi}\int_{r_b}^{R}\!(r^2 - n^2)dr 
\int_{0}^{\pi}\!\int_{0}^{2\pi}\!\sin{\theta}d{\theta}d{\varphi}}
and
\eqn\morexplcit{{\rm Vol}_{n}(R) =
 2n\int_{0}^{4\pi}\! d{\psi}\int_{n}^{R}\! (r^2 - n^2)dr
\int_{0}^{\pi}\!\int_{0}^{2\pi}\!\sin{\theta}d{\theta}d{\varphi}}
so that the total volume difference is given as the limit, as
$R{\rightarrow}\infty$, of ${\rm Vol}_{b}(R){-}{\rm Vol}_{n}(R)$. 

Recalling that we must ensure that the induced metrics of TN--AdS and
TB--AdS match on the hypersurface ($r{=}R$), we see that we must
rescale the nut parameter $n_{n}$ of TN--AdS to $\lambda(r)n_{b}$
(where $n_{b}$ is the nut parameter of TB--AdS), in order that their
Euclidean times have the same period to sufficiently high order. (The
function $\lambda(r)^2$ is obtained by expanding the ratio of the
metric functions $V(r,n,m)$ obtained in each case.) 

In this way we find
\eqn\found{n_{b} = n_{n} \left( 1+ {l^2 (m_{b}- 
m_{n}) \over R^3} +O(R^{-4})
\right).}
Putting all of this together one therefore obtains the final result
for the action of TB--AdS after considerable algebra:
\eqn\final{I_{b} 
= -{2\pi n\over G l^2}\left({(r_b-n)^2 (r_b^2-6nr_b-n^2)\over
r_b-2n}\right).}  We can now analyze for which values of the nut
parameter $n$ the action of TB--AdS is larger or smaller than that of
TN--AdS, ie, where $I_{b}$ is positive or negative.  A short inspection
shows that $I_{b}$ is positive only in the range
$2n{<}r_b{<}(3{+}\sqrt{10})n$ (of course, we are always considering
$r_b{\geq}n$).  Figure~1 is a plot of $r{=}r_b$ as a function of $n$,
in the allowed range of variables $r_b{<}n_{max}$. We also include the
lines $r{=}2n$ (dotted), and $r{=}(3{+}\sqrt{10})n$ (dashed).

\midinsert{
\centerline{\epsfxsize3.0in\epsfbox{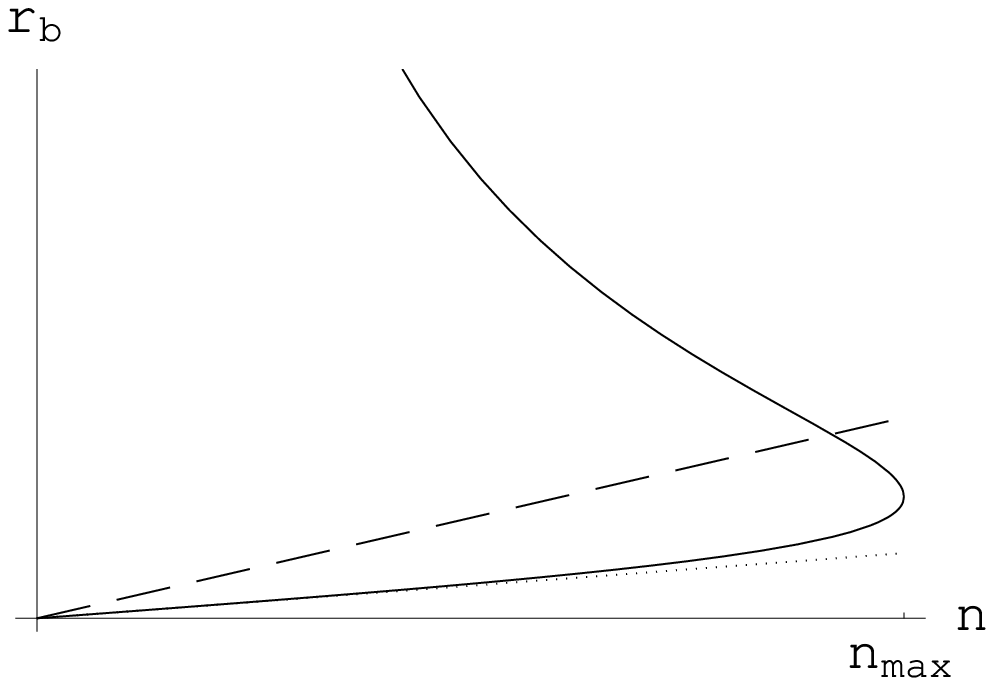}}
{\sl Fig 1: Plot of $r_b$ as a function of $n$, up to $n_{\rm max}$
which limits the existence of TB--AdS solutions.  The straight lines
are $r_b{=}2n$ (dotted) and $r_b{=}(3{+}\sqrt{10})n$ (dashed). }
}\endinsert

We can see that we always have $r_b{>}2n$ from \boltplace. Note that
$r_{b+}{\to}2n$ as $n{\to}0$.  The lower branch $r_{b-}$ lies entirely
between $r{=}2n$ and $r{=}(3{+}\sqrt{10})n$, and so the action is always
positive for these solutions. On the upper branch $r_{b+}$, the action
is positive for the smallest values of $r_{b+}$ (the largest values of
$n$), but as $r_{b+}$ grows ($n$ becomes smaller), the action becomes
negative. The crossover point, {\it i.e.} $I_b{=}0$, lies at
$n{=}n_{\rm crit}{=}l(7{-}2\sqrt{10})^{1/2}/6$.

\newsec{Some Thermodynamics}
We have performed a covariant computation of the action, as distinct
from a Hamiltonian calculation, which would have required a specific
time slicing. Such a calculation would have identified a periodic time
in an ADM manner\adm, using the temperature $T{=}1/(8\pi n)$. We
expect that such a calculation would have shown that the action
decomposes into contributions from the Hamiltonian of the Misner
strings at infinity, in addition to the usual terms corresponding to
the area of the bolt\gibbhawk. 

We will not carry out a Hamiltonian calculation here, instead moving
on to compute various state functions and hence study the physics of
the present situation.

We have for the entropy the formula $S=(\beta \partial_\beta{-}1) I.$
 Lengthy algebraic manipulations finally yield the entropy in a simple
 form
\eqn\entropy{S={\pi \over G} (r_b-n)^2 \left(1+12 {n^2\over l^2}\right)}
This is manifestly positive. It should be noted that this expression 
differs from $A_{bolt}/4G$: there are contributions to the entropy 
from the nut charge and nut potential at the bolt\gibbhawk.

We plot the entropy as a function of $n$ in figure~2, including that
of the lower branch solutions.

\midinsert{
\centerline{\epsfxsize3.0in\epsfbox{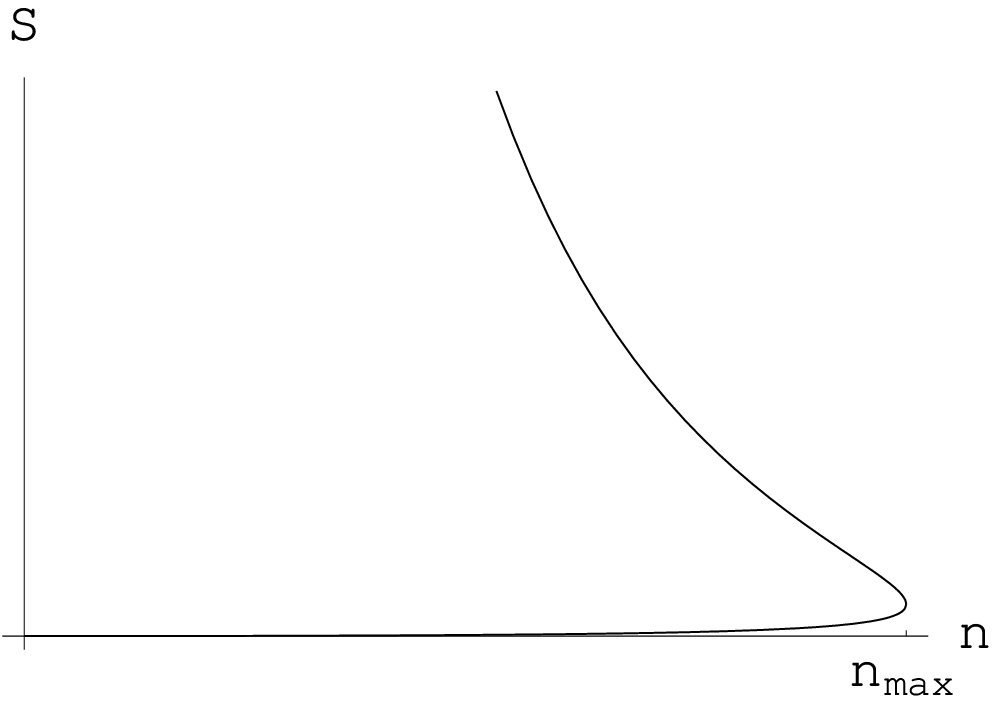}}
{\sl Fig 2: The entropy, $S$, as a function of $n$.}
}\endinsert

We can compute the thermodynamic energy $E$,
\eqn\energy{E=
\partial_\beta I={1 \over 2Gl^2} {(r_b-n)^3 (r_b+7n) 
\over r_b-2n}= {m_{b}-m_{n}\over G}}
where $m_{b,n}$ are the mass parameters for TB--AdS and TN--AdS as
given in equations \massone\ and \implies\ above.  Since $r_b{>}2n$,
the energy is strictly positive.

We are particularly interested in the very high temperature regime, 
$n{\rightarrow}0$. In this limit we have
\eqn\higharby{r_{b+}= {l^2 \over 6n} - 2n +O(n^3).}
For the upper branch solutions, the action and entropy become
\eqn\highaction{I=-{\pi l^4 
\over 108 G n^2}+O(n^0) \qquad S={\pi l^4 \over 36 G 
n^2}+O(n^0)} The entropy coincides in this limit with the limiting
value of $A_{bolt}/4G$, showing that in the high temperature regime
the effect of the non--trivial topological fibering of the manifold
(the contribution from the Misner string\misner) becomes invisible, as
could be expected.

Note that the lower branch solutions (which have higher action and
lower entropy) have the following behaviour at high temperature in the
limit $\Lambda{\to}0$:
\eqn\behave{r_b = 2n +O(n^3); \quad I_b={\pi n^2\over G} + O(n^4);\quad
S = {\pi n^2\over G} +O(n^4).} These are the values obtained in the
$\lambda{=}0$ Taub--NUT/Bolt action calculations of ref.\hunt.  This
is entirely consistent with the observation, made in section~2.2, that
the lower branch bolt solutions tend to the $\Lambda{=}0$ solutions in
this limit.

Focussing on the upper branch solutions (which will always be more
stable, see later), we immediately see\foot{Crucially, use the fact
that this is an eleven dimensional supergravity compactification, so
$G{\sim}l^{-7}$ (in units where the eleven dimensional Planck length
is unity) and $l{\sim}N^{1/6}$}\ that the free energy
$F{\sim}V_2T^3N^{3/2}$ and entropy $S{\sim}V_2T^{2}N^{3/2}$, ($V_2$ is
the spatial volume of the field theory) which corresponds to the
expected high temperature behaviour of a field theory in three
space--time dimensions. It is important to note that the growth with
$N$ is slower than $N^2$, confirming that the $N$ is {\sl not}
associated with the gauge theory of $N$ D2--branes in ten spacetime
dimensions, but rather the more exotic field theory associated to $N$
M2--branes in eleven dimensions. (The former flows to the latter in
the infra--red\seiberg.) The power $N^{3/2}$ counts the number of
degrees of freedom of the theory, showing that we are in, roughly
speaking, a deconfined phase of the theory. The $N^{3/2}$ factor was
first noted in ref.\igor\ as associated with the entropy of $N$
coincident M2--branes. We consider our present calculations, with
their holographic interpretation, as independent support for the
conclusion of ref.\igor\ that the 2+1 dimensional CFT has O($N^{3/2}$)
degrees of freedom. (This also follows from the results of
refs.\refs{\ed,\edd}\ for the AdS$_4$/AdS$_4$--Schwarzschild case,
once the appropriate conversions have been made.)

Recall that the Taub--bolt--AdS solutions only existed for
$n{<}n_{\rm max}$; the radius of the bolt becomes unphysical.  This
means that below a certain temperature $T_{\rm min}{=}1/(8\pi n_{\rm
max})$, the solution does not exist, and the TN--AdS solution is the
allowed one. Above that temperature, there is apparently a transition
to the TB--AdS solution as is evident from the displayed plots in
figs.~2 and~3. However, the transition at $T_{\rm min}$ is merely an
artifact of the fact that we not truly in the thermodynamic
limit. More careful consideration below will reveal the transition to
be at a higher temperature, $T_{\rm crit}$.

In order to study the thermal stability of the system it is convenient
to examine the specific heat $C=-\beta \partial_\beta S = -n
\partial_n S$. The analytical expression, however, is not very
illuminating. Instead, we provide in figure~3 a plot of $C$ as a
function of $n$ which remains positive for the upper branch solutions,
negative for the lower branch solutions, and begins to grow rapidly
near $T_{\rm min}$ for both branches.

\midinsert{
\centerline{\epsfxsize3.0in\epsfbox{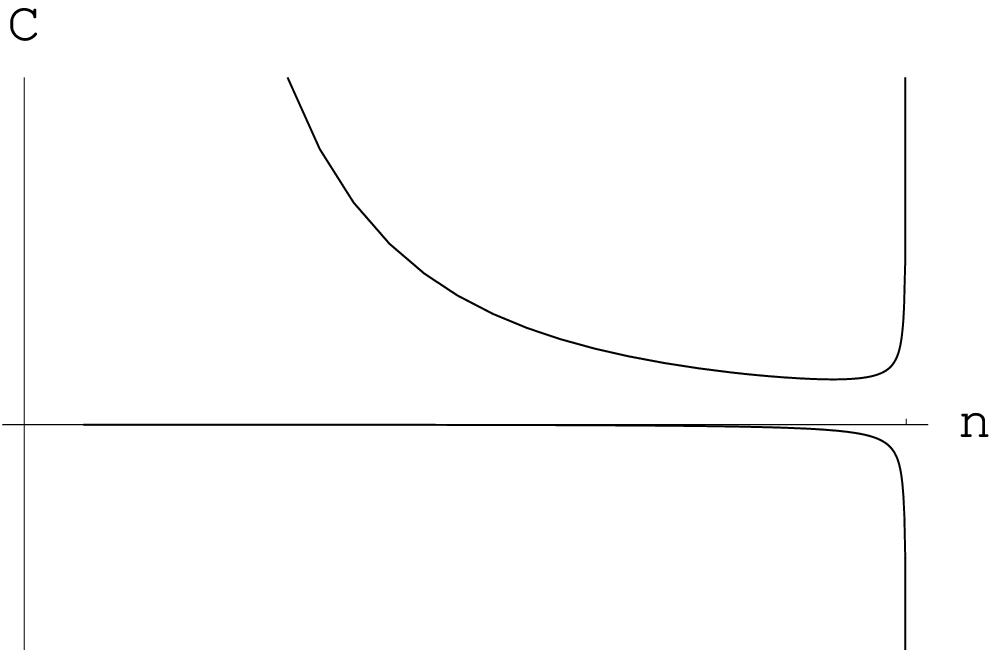}}
}{\sl Fig. 3. Specific heat, $C$, at constant volume, as a function
 of $n$. It diverges sharply at $n{=}n_{\rm max}$.}
\endinsert

Notice however that the action \final\ is {\sl positive} for $T_{\rm
min}{<}T{<}T_{\rm crit}{=}1/(8\pi n_{\rm crit})$. Above $T_{\rm crit}$ it
goes negative. This means that the TN--AdS solution is still relevant above
$T_{\rm min}$. As the specific heat of TB--AdS is positive however, we
can have a stable bolt as well, and therefore we can nucleate
long--lived bolt solutions.

This initial conclusion that there is a  phase transition
to a (nearly) co--existence phase is modified by the fact that we have
not taken the limit $N{\to}\infty$. To see how the modification comes
about, first recall that (after converting $G{\sim}l^{-7}$ and
$l{\sim}N^{1/6}$) there is a positive power of $N$ multiplying the
action $I_b$, and remember that $I_b$ is the {\it difference} between
that of AdS--TB and AdS--TN.

Therefore in the thermodynamic limit where we take the number of
degrees of freedom (measured by $N$) to infinity, the action
difference is {\it infinite}. We must conclude therefore that the true
phase transition takes place where the {\sl sign} of $I_b$ changes,
which is at $T_{\rm crit}$. The free energy is continuous there, but
the energy $E$ is discontinuous, and so we conclude that it is a {\sl
first order} phase transition: The degrees of freedom are distinct in
each case, as shown by the fact that the amount of entropy associated
with thermal radiation in TN--AdS is vastly exceeded by the amount
which can be stored in the area of the bolt (and Misner strings) in
the TB--AdS solution.

(Notice that this analysis and discussion also applies to the
AdS$_4$/AdS$_4$--Schwarzschild case studied in ref.\refs{\ed,\edd},
although the more complex thermodynamic conclusions made about the
same bulk physics in ref.\hawkpage\ are more general, as they are not
restricted to the large~$N$ limit of this context.)

We remark again that although this represents the physics of
transitions between very different gravitational solutions, the
complete physics is very plausibly described by  the unitary
conformal field theory living on the twisted three sphere at the
boundary.

\newsec{Topologically Trivial Nuts and Bolts}

The Taub--NUT--AdS family of metrics contains solutions where the 
angular spheres $(\theta,\varphi)$ are replaced by planes, or 
hyperboloids. For vanishing nut charge, the solutions correspond to 
topological black holes\topo, studied in ref.\roberto\ in their 
M--theory context.

\subsec{Planar Nuts and Bolts}

Let us focus first on the planar (or toroidal) solutions
\eqn\toroidal{ds^2 = V(r) \left(d\tau + {n\over l^2} (xdy-ydx)\right)^2 +
 {dr^2\over V(r)} + {r^2 -n^2 \over l^2}(dx^2 +dy^2)} where, now,
\eqn\now{V = { - 2 mr  + l^{-2} (r^4 - 6 n^2 r^2-3 
n^4)\over r^2 - n^2}.}  The coordinates $x$, $y$ here have dimensions
of length. Notice that the fibration is now trivial: there are no
Misner strings. The topology of the boundary at $r\rightarrow\infty$
is therefore $\IR^3$. However, although the boundary is topologically
a direct product of the Euclidean time line, and the spatial plane
$(x,y)$, the product is ``twisted'' or warped, and the boundary is
not flat.

An immediate consequence of the trivial topology is that the Euclidean
time period $\beta$ will not be fixed, as it was in the spherical
case, by the value of the nut parameter\foot{Note that if $\tau$, $x$
and $y$ are all compactified on a (warped) torus $T^3$, consistency
will demand that the period $\beta$ is fixed in terms of $n$. We will
not do such a compactification here.}\ $n$. Therefore, in the present
case we can vary the temperature of the system while leaving $n$
fixed. In other words, $n$ labels different sectors of the theory,
characterizing the ``warpage'' of the product $\IR{\times}\IR^2$. For
each sector, we can consider the phase structure as a function of
temperature separately.

In the absence of Misner strings, we expect the entropy of the 
solutions to receive contributions solely from the area of bolts. 
This expectation will be confirmed below.

Let us now proceed to examine the fixed-point sets of the isometry 
generated by $\partial_\tau$ ---the planar nuts and bolts. Nuts will 
appear as fixed-point sets at $r=n$. One finds that the mass 
parameter must take the value
\eqn\value{m_{n}=-{4n^3 \over l^2}}
so that
\eqn\sothat{V_{n}(r)={(r-n)^2 (r+3n) \over l^2 (r+n)}.}
Notice that $V_{n}(r)$ has a double zero at $r{=}n$. This is, the 
solution must be regarded as an extremal, zero temperature 
background, since the Euclidean time $\tau$ can be identified with 
arbitrary period. In fact, when $n{=}0$ we simply recover the AdS$_4$ 
metric in horospheric coordinates $(r{=}1/z)$. 
It is also interesting to note that the mass parameter is negative 
for all other cases. This might be an indication that the CFT defined 
on these boundary geometries might be unstable. Although we have not 
checked this point, these backgrounds are presumably 
non--supersymmetric for $n{\neq}0$.

Now let us find Taub--bolt--AdS solutions, where $\partial_\tau$ has a
two dimensional fixed--point set at some radius $r{=}r_b{>}n$. In this
case we find that the mass parameter has to be
\eqn\mass{m_{b}=
{1\over 2l^2} \left( r_b^3 - 6n^2 r_b - {3 n^4 \over r_b} 
\right).}
This time, Euclidean regularity at the bolt requires the period of 
$\tau$ to be 
\eqn\tobe{\beta={4\pi \over V'(r_b)}= {4\pi l^2 \over 3} {r_b \over r_b^2 
-n^2}.}  As $r_b$ varies from $n$ to infinity, we cover the whole
temperature range from $0$ to $\infty$. Notice that $m_{b}$ can be
either negative, zero, or positive. When $n{=}0$ we recover the standard
results for Schwarzschild--AdS$_4$.

As we said above, we can thermally excite each of the sectors 
labeled by $n$, keeping $n$ fixed. This requires us to study the 
thermodynamics of TB--AdS solutions above a TN-AdS background with the 
same nut charge. As usual, in order to match the geometries at large 
radius $R$ we must set
\eqn\set{\beta_n\sqrt{V_n(R)}=\beta_b\sqrt{V_b(R)}.}
We must also match the values of the nut charges, but this turns out 
to yield a contribution to the action that vanishes as $R{\rightarrow} 
\infty$, and therefore will be neglected. The computation of the 
action, which is reduced to a difference of volume terms, is 
straightforward, and yields
\eqn\yield{I_b=-{L^2 \over 12 G 
l^2}\left({r_b-n\over r_b+n} (r_b^2 +2 nr_b +3n^2)\right)} where $L^2$
accounts for the area of the $(x,y)$ plane, $-L/2{\leq}x, y{\leq}L/2$.

Now we find
\eqn\energy{E={L^2 \over 8\pi G l^4}(r_b-n)^2 (r_b+2n).}
Notice that, for $n\neq 0$, this is different from the value
\eqn\diff{{L^2 \over 4\pi G l^2} (m_b-m_n)= {L^2 \over
 8\pi G l^4}(r_b-n)^3 (r_b+3n)} that could, perhaps, have been
expected. This means that in this case one should not think of $m$ as
a parameter directly related to the mass.

The action is, for $r_b{>}n$, always negative. Therefore, like in the 
$n{=}0$ case, there are no phase transitions as a function of the 
temperature and the system stays always in the ``deconfined'' phase.

Finally, the entropy
\eqn\entropytwo{S=
(\beta \partial_\beta -1)I={L^2 (r_b^2-n^2)\over 4 G l^2} = {A_{bolt}
\over 4G}} reproduces the Bekenstein--Hawking\beckhawk\ result, as it
should in the absence of Misner strings.

At high temperatures the entropy behaves in the conformally invariant 
way $S{\sim}\beta ^{-2}$. In this regime, the non--trivial warpage for 
$n{\neq}0$ is invisible. However, at lower temperatures the entropy 
departs from the CFT behavior. This is as expected, since the warpage 
breaks conformal invariance by introducing a non--vanishing scale, 
namely, the mass parameter~$m$.

\subsec{No Hyperbolic Nuts}

There is also the possibility of having hyperbolic, instead of 
spheric or planar, fixed--point sets of $\partial_\tau$. The metric to 
be used is, in this case,
\eqn\case{ds^2 =
 V(r) (d\tau + 2n (\cosh\theta-1) d\varphi)^2 + {dr^2\over V(r)} +
(r^2 -n^2)(d\theta^2 +\sinh^2\theta d\varphi^2)} with
\eqn\with{V = { -(r^2 +n^2)- 2 mr  +l^{-2} (r^4 - 6 n^2 r^2-3 
n^4)\over r^2 - n^2}.}
The coordinates $(\theta,\varphi)$ parametrize a hyperboloid, and 
upon appropriate quotients, surfaces of any genus higher than $1$. 
The fibration is trivial, and, again, there are no Misner strings.

However, if we try to make $r{=}n$ into a fixed point of  
$\partial_\tau$, we find that $V(r)$ becomes negative for $r$ close 
enough to (and bigger than) $n$. That is, $V$ vanishes at some $r{>}n$, 
and instead of a nut we find a bolt. Thus, there are no hyperbolic 
nuts.

One could study the thermodynamics of these solutions by taking as a 
background a singular, extremal bolt. However, the holographic 
significance of these solutions is obscure, as it is for $n=0$, where 
it has been argued that these systems are likely to be unstable\roberto.

\newsec{Conclusions}
Having proposed that it should be instructive to revisit the program
of studying various quantum gravity processes in the light of the
holographic principle (as embodied by the use of AdS), we have
enlarged the arena somewhat by studying some examples which are only
{\it locally} asymptotically AdS.

The boundary conformal field theory is the Euclideanized $2+1$
dimensional superconformal field theory compactified on a squashed
three sphere, in one case, and a twisted plane in another, mapping its
phase structure to that of  Taub--NUT--AdS/Taub--Bolt--AdS systems
in the bulk.

We find that at high and (to a lesser extent) low temperatures, the
thermodynamic properties of the theory are those we expect from
general considerations, and are consistent with the properties of the
dual conformal field theory, including an unambiguous phase transition
at $T_{\rm crit}$. It would be interesting to study further the
properties of the field theory at intermediate and low temperatures.

In section 2.3 we suggested that AdS$_4$ could not be used as a
background solution in the action calculations. This is because we
were unable to embed the asymptotic squashed $S^3$ into AdS$_4$. If
this were possible, the phase structure could be even more complicated
by the introduction of an AdS$_4$ phase. However, the fact that our
results are consistent with the field theory equivalence suggests that
our calculations are correct without such a contribution (at least at
high temperatures).

We now have three concrete families of holographic examples of the map
between the large $N$, finite temperature properties of 2+1 dimensional
field theory and gravity: that of refs.\refs{\ed,\edd}\ and those
presented here.

As we stressed in the introduction, this program has sharpened the
debate about the nature of various instanton calculations in quantum
gravity, and may provide a practical answer to the question of the
unitarity of processes which including space--like topology change.
We intend to report on further examples in the near future.

\bigskip
\bigskip

\rightline{\vbox{\hbox{``Don't worry if your lot is small}
\hbox{And your rewards are few;}
\hbox{Remember that the Mighty Oak}
\hbox{Was once a {\sl nut} like you!''}
\hbox{\it ---Anonymous (possibly B.M.)}}}

\bigskip
\bigskip

\vfill\eject

{\noindent \bf Acknowledgments}

AC would like to thank Stephen Hawking and Chris Hunter for useful
conversations. CVJ and RCM would like to thank Miao Li for a useful
conversation. AC is supported by Pembroke College, Cambridge.  RE is
supported by EPSRC through grant GR/L38158 (UK), and by grant UPV
063.310--EB225/95 (Spain).  CVJ's research was supported by an NSF
Career grant, \# PHY97--33173 (UK), and NSF grant \# PHY97--22022
(UCSB).  RCM's research was supported by NSERC (Canada), Fonds FCAR du
Qu\'ebec, and NSF grant \# PHY94--07194 (ITP, UCSB). CVJ thanks the
members and staff of the Abdus Salam International Center for
Theoretical Physics at Trieste, the organizers and participants of the
1998 Trieste Spring School on ``Non--Perturbative Aspects of String
Theory and Supersymmetric Gauge Theory'' for a stimulating atmosphere,
and the Physics Department of Columbia University for hospitality. CVJ
and RCM would like to thank the staff of the Aspen Center for Physics
and the participants and organizers of the program ``M--Theory and
Black Holes'', for another stimulating atmosphere for research. We all
thank the members and staff of the Institute for Theoretical Physics
and the Physics Department, UCSB, and the participants and organizers
of the ``String Duality'' workshop and the ``Strings '98'' conference
for atmospheres no less stimulating than those already mentioned.

\listrefs 
\bye